\documentstyle[epsfig,12pt]{article}
\newcommand{\ur}[1]{(\ref{#1})}
\newcommand{\urs}[2]{(\ref{#1},\ref{#2})}

\newcommand{\eq}[1]{eq.~(\ref{#1})}
\newcommand{\eqs}[2]{eqs.~(\ref{#1}, \ref{#2})}
\newcommand{\Eq}[1]{Eq.~(\ref{#1})}

\def\Tr{\mbox{Tr}}
\newcommand{\la}[1]{\label{#1}}
\def\beq{\begin{equation}}
\def\eeq{\end{equation}}
\def\bea{\begin{eqnarray}}
\def\eea{\end{eqnarray}}

\begin{document}
\thispagestyle{empty}
\begin{flushright} NORDITA-1999/47 HE
\end{flushright}
\vskip 2true cm
\begin{center}
{\Large\bf Vortex Solution in 2+1 Dimensional Pure Yang--Mills \\
\vskip .5true cm
Theory at High Temperatures} \\
\vskip 1.5true cm

{\large\bf Dmitri Diakonov}
\\[1cm]
{\it NORDITA, Blegdamsvej 17, 2100 Copenhagen \O, Denmark}\\
and\\
{\it Petersburg Nuclear Physics Institute, Gatchina,
St.Petersburg 188350, Russia} \\
\vskip .5true cm
E-mail: diakonov@nordita.dk
\end{center}
\vskip 1.5true cm
\begin{abstract}
\noindent At high temperatures the $A_0$ component of the Yang--Mills
field plays the role of the Higgs field, and the 1-loop potential
$V(A_0)$ plays the role of the Higgs potential. We find a new stable
vortex solution of the Abrikosov--Nielsen--Olesen type, and discuss
its properties and possible implications.
\end{abstract}

\vskip 2true cm

Recently there has been renewed interest in the quantized $Z(N_c)$
vortices as possible candidates for the confinement mechanism
\cite{G1,G2,G3}. If vortices are physical objects (and not lattice or
gauge artifacts) playing a role in the dynamics of the vacuum
fluctuations they have to be found as stable solutions of the effective
action obtained from integrating out high frequencies of the field. In
the zero-temperature case the one-loop effective action for vortices
has been introduced in our previous work \cite{D}. The zero-derivative
term of the effective action, i.e. the `potential energy' of the vortex
has been found in that paper: it does not lead to any stable solution.
This is probably not surprising since all derivatives of the effective
action should be summed up before one gets to a definite conclusion on
the existence of vortex solutions.

In this letter we address a related but simpler question on whether
there are stable vortex solutions in the case of {\em nonzero
temperatures}. Moreover, we restrict ourselves to a simple and more `pure'
case of $2+1$ dimensions. If the temperature $T$ is high enough
vortices can be viewed as cylinders pointing in the `time' direction,
with a nontrivial profile in the transverse `spatial' plane. We shall
show that the Abrikosov--Nielsen--Olesen-type vortex solution indeed
exists in this case, and we shall present its profile and energy per unit
length, as functions of temperature.

The $2+1$-dimensional YM theory has a coupling constant $g_3^2$ of the
dimension of mass; high temperatures means $T\gg g_3^2$. One can always
choose a gauge with the `time' component of the YM field $A_0^a(x)$
independent of `time'. In this gauge, one can integrate out the
high-momenta components of the fields to obtain the low-momenta
effective action for time-independent fields $A_\mu^a(x),\;\mu=0,1,2$.
The separation scale between high and low momenta is given, naturally,
by the temperature: high momenta means $p>2\pi T$. The effective action
is obtained by integrating over nonzero Matsubara frequencies,
$\omega_n=2\pi n T,\;n>0$, and over zero ($n=0$) Matsubara frequency but
large spatial momenta. It can be expanded in powers of the spatial
derivatives of the $A_\mu^a(x)$ fields, divided by appropriate powers of
$T$.  The zero-derivative term is the `potential energy' $V(A_0)$. Since
the spatial size of the vortex solution is expected to be larger than
$1/T$ one can neglect all terms of the derivative expansion except the
first, zero-derivative term $V(A_0)$. The latter can be written as
\cite{KA,D}

\bea
V(A_0)&=&\frac{2T^3}{\pi}\sin^2\pi\nu\int_0^\infty\!dp
\,\frac{p^2\cosh p}{\sinh p\,(\sinh^2 p+\sin^2\pi\nu)}
=\pi T^3\nu^2\left(\ln\frac{1}{\nu^2}+2.67575\right)+O(\nu^4)
\nonumber\\
&=&\frac{T}{4\pi}(A_0^a)^2\ln\frac{{\rm const}\cdot T^2}{(A_0^a)^2}
+O(A_0^4), \qquad\nu=\frac{\sqrt{A_0^aA_0^a}}{2\pi T}.
\la{potA0}\eea
It is a periodic function of the dimensionless variable $\nu$ with unit
period, depicted in Fig.1.  Notice that the potential energy
\ur{potA0} is nonanalytic at the minima $\nu\,=\,integer$. The fact that
the second derivative of $V(A_0)$ has a logarithmic singularity at
$\nu\,=\,integer$ is related to the infrared divergency of the Debye mass
in $2+1$ dimensions \cite{DH,Zhuk}, and will have important consequences
for the vortex solution.

\begin{figure}
\centerline{\epsfxsize10.0cm\epsffile{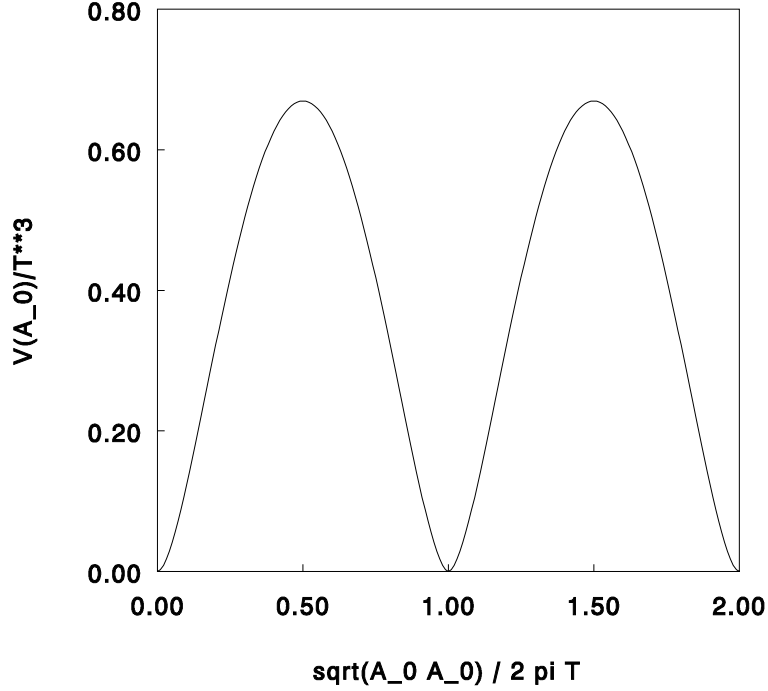}}
\caption[]{{\em $2+1$-dimensional periodic potential $V(A_0)$,
\eq{potA0}.}}
\end{figure}

Adding $V(A_0)$ to the classical YM action,
$F_{\mu\nu}^aF_{\mu\nu}^a/4g_3^2$, we get the energy functional in the
spatial transverse plane:

\beq
E_\perp=\int d^2x\left\{\frac{1}{2g_3^2}\left[\epsilon_{ij}
\left(\partial_i A_j^a+\frac{1}{2}f^{abc}A_i^bA_j^c\right)\right]^2
+\frac{1}{2g_3^2}\left(D_i^{ab}A_0^b\right)^2+V(A_0^a)\right\}
\la{Eperp1}\eeq
where $i,j=1,2$ denote spatial components,
$D_i$ is the covariant derivative,
$D_i^{ab} = \partial_i\delta^{ab}+f^{acb} A_i^c$.

\Eq{Eperp1} presents a 2-dimensional YM $+$ Higgs system where $A_0^a$
plays the role of the adjoint Higgs field, with a peculiar form of the
potential energy \ur{potA0}.

We choose the following vortex-type Ansatz for the fields, restricting
ourselves to the $SU(2)$ colour group for simplicity:

\bea
A_i^a&=&\delta^{a3}\epsilon_{ij}n_j\,\frac{\mu(\rho)}{\rho},\qquad
\rho=\sqrt{x_1^2+x_2^2}, \qquad n_j=\frac{x_j}{\rho},
\la{Ai}\\
A_0^a&=&\delta^{ai}\epsilon_{ij}n_j \,\nu(\rho)\,2\pi T,
\la{A0}\eea
where $\mu(\rho)$ and $\nu(\rho)$ are trial profile functions of the
distance $\rho$ from the vortex center. The Ansatz \ur{Ai}
corresponds to the radial component $A_\rho^a=0$ and the azimuthal
component $A_\phi^a(\rho)=\delta^{a3}\mu(\rho)/\rho$. The closed Wilson
loop in the representation labelled by spin $J$, circling around the
center of the vortex in the transverse plane at distance $\rho$, is

\beq
W_J(\rho)=
\frac{1}{2J+1}\Tr\;{\rm P}\exp i \oint d\phi\rho A_\phi^a T^a=
\frac{1}{2J+1}\frac{\sin[(2J+1)\pi\mu(\rho)]}{\sin[\pi\mu(\rho)]},
\qquad \mu(\rho)=\sqrt{A_\phi^a A_\phi^a}\rho,
\la{WJ}\eeq
where $\mu(\rho)$ is the magnetic field flux along the vortex inside a
tube of radius $\rho$. If $\mu(\rho)\to 1$ at large $\rho$ the Wilson
loop $W_J(\rho)\to (-1)^{2J}$. In particular, in the fundamental
representation one has $W_{1/2}\to -1$. This is the definition of the
quantized $Z(2)$ vortex. Our solution will be precisely of this type.

Let us introduce the dimensionless radius

\beq
x=\rho g_3^{1/2} T^{3/4}
\la{x}\eeq
and the dimensionless parameter

\beq
\alpha=\frac{g_3}{\surd T (2\pi)^2}.
\la{alpha}\eeq
At high temperatures $\alpha \ll 1$. In terms of these quantities the
action functional for the vortex of length $1/T$ becomes

\beq
S_{{\rm vortex}}=\frac{E_\perp}{T}
=\frac{1}{2\pi\alpha^2}\int_0^\infty\! dx\, x
\left[\frac{1}{2}\nu^{\prime\,2}+\frac{1}{2x^2}\nu^2(1-\mu)^2
+\alpha\frac{1}{2x^2}\mu^{\prime\,2}+\alpha v(\nu)\right],
\la{Eperp2}\eeq
where we have introduced the dimensionless `Higgs potential'

\beq
v(\nu)=\frac{2}{\pi}\sin^2\pi\nu\!\int_0^\infty\!\! dp\,
\frac{p^2\cosh p}{\sinh p\,(\sinh^2 p+\sin^2\pi\nu)}=
\left\{\begin{array}{cc}\pi\nu^2\ln\frac{1}{\nu^2}+...
&{\rm at}\;\nu\approx 0, \\
\pi(1-\nu)^2\ln\frac{1}{(1-\nu)^2}+...&{\rm at}\;\nu\approx 1.
\end{array}\right.
\la{reducedv}\eeq
The Euler--Lagrange equations of motion for the profile functions
$\mu,\nu(\rho)$ are:

\bea
\alpha\frac{d}{dx}\left(\frac{1}{x}\frac{d\mu}{dx}\right)&=&
-\frac{1}{x}\nu^2(1-\mu),
\la{eqmotmu}\\
\frac{d}{dx}\left(x\frac{d\nu}{dx}\right)&=&
\frac{1}{x}\nu(1-\mu)^2+\alpha x\frac{dv}{d\nu},
\la{eqmotnu}
\eea

\beq
\frac{dv}{d\nu}=2\sin 2\pi\nu\!\int_0^\infty\!\frac{dp\, p}
{\sinh^2 p +\sin^2\pi\nu}
=\left\{\begin{array}{cc}2\pi\nu\left(\ln\frac{1}{\nu^2}+1.67575\right)
&{\rm at}\;\nu\approx 0, \\
-2\pi(1-\nu)\left(\ln\frac{1}{(1-\nu)^2}+1.67575\right)
&{\rm at}\;\nu\approx 1.
\end{array}\right.
\la{dvpodnu}\eeq
We look for the solution with the following boundary conditions:

\beq
\mu(0)=\nu(0)=0,\qquad\mu(\infty)=\nu(\infty)=1,
\la{boundary}\eeq
corresponding to the quantized $Z(2)$ vortex, with the `Higgs field'
$A_0$ going from the trivial minimum at $\rho=0$ to a non-trivial
one at $\rho\to\infty$.

The behaviour of the profile functions near the origin and at infinity
can be found analytically. At small $x$ we get:

\bea
\mu(x)&=&c_1x^2+c_2x^4+...\;,\nonumber\\
\nu(x)&=&d_1x+d_2x^3\ln x+d_3x^3+...\;.
\la{smallx}\eea
The coefficients $c_1,d_1$ are arbitrary but the higher
coefficients are determined from \eqs{eqmotmu}{eqmotnu}:

\beq
c_2=-\frac{d_1^2}{8\alpha},\quad d_2=-\frac{\alpha\pi d_1}{2},\quad
d_3=\frac{d_1}{8}\left[-2c_1+\alpha\pi(3+2h-4\ln d_1)\right],\quad
h=1.67575...
\la{constants0}\eeq
At large values of $x$ the analytical solution is:

\bea
\mu(x)&\approx&1-e_1\sqrt{x}\exp\left(-\frac{x}{\sqrt{\alpha}}\right),
\la{largexmu}\\
\nu(x)&\approx&1-f_1\exp\left(-\pi\alpha x^2\right),
\la{largexnu}\eea
where the constants $e_1,f_1$ are not determined by the equations.
Notice that the `Higgs field' $\nu(x)$ approaches its asymptotic value
at infinity not as an exponent but as a gaussian. This is a
consequence of the logarithmic divergence of the Debye mass in $2+1$
dimensions. Being rewritten in original notations
\eq{largexnu} reads

\beq
\left|A_0^a(\rho)\right| \rightarrow
2\pi T\left[1-f_1\exp\left(-\frac{g_3^2T}{4\pi}\rho^2\right)\right],
\qquad {\rm at}\;\rho\to\infty.
\la{largerhoA0}\eeq
We notice that the characteristic scale for the variation of
$A_0(\rho)$, namely $1/g_3\surd T$, is, at high temperatures, much
larger than $1/T$. It justifies neglecting derivative terms of $A_0$
in the effective action and leaving only the zero-derivative term
$V(A_0)$, as in \eq{Eperp1}.

The coefficients $c_1,d_1,e_1,f_1$ are determined by solving
\eqs{eqmotmu}{eqmotnu} starting from the expansion \ur{smallx}
towards larger values of $x$ and starting from the asymptotic form
\urs{largexmu}{largexnu} towards smaller values of $x$, and matching
the functions and their derivatives at some intermediate point
$x\sim 1$; this is done numerically. The resulting profile functions
$\mu(x)$ and $\nu(x)$ are shown in Fig.2, for two values of
$\alpha=0.15$ and 0.5.

\begin{figure}
\centerline{\epsfxsize10.0cm\epsffile{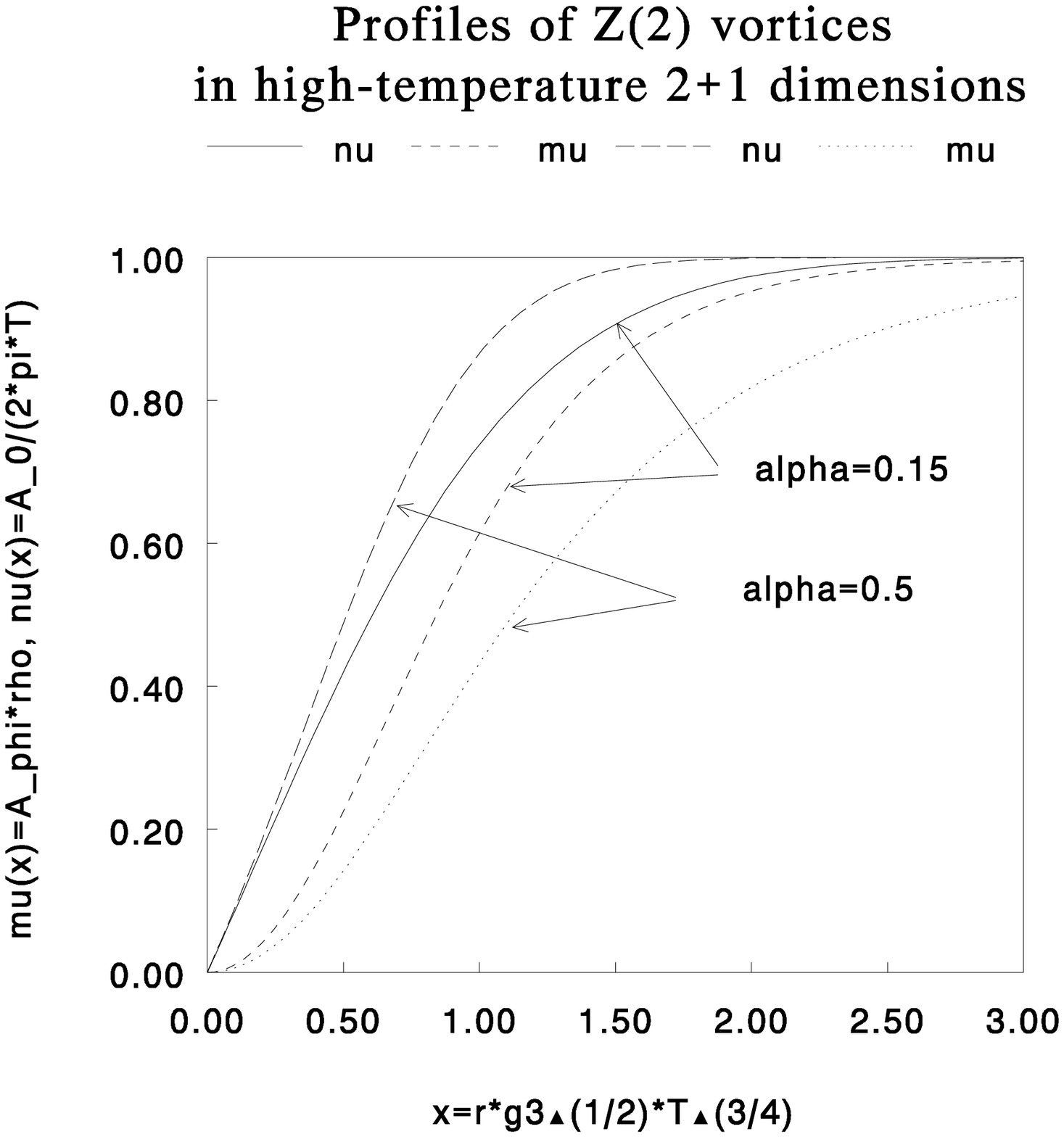}}
\caption[]{{\em Profile functions of the vortex for two values of
temperature, corresponding to $\alpha=$0.15 and 0.5.}}
\end{figure}

Integrating the profile functions over the whole range of $x$ we get
the action of the vortex $S_{{\rm vortex}}=E_\perp/T$
plotted in Fig.3, as function of temperature $T$.
At $T\gg g_3^2$ the action is large, so that the vortices are
exponentially suppressed. This is a theoretically `clean' case:
as explained above, the use of the energy functional \ur{Eperp1}
for finding the vortex solution is justified. The vortices are
in fact short and broad cylinders oriented in the `time' direction.
Indeed, the length of the cylinders is $1/T$ while the characteristic
radius where the `Higgs field' $A_0$ reaches its asymptotic value
$2\pi T$ is of the order of $1/g_3\surd T\gg 1/T$, see \eq{largexnu}.
It should be mentioned, however, that the magnetic flux
$\mu(\rho)$ reaches its asymptotic value of 1 at a smaller radius
$\rho\sim 1/T$, see \eq{largexmu}.

\begin{figure}
\centerline{\epsfxsize10.0cm\epsffile{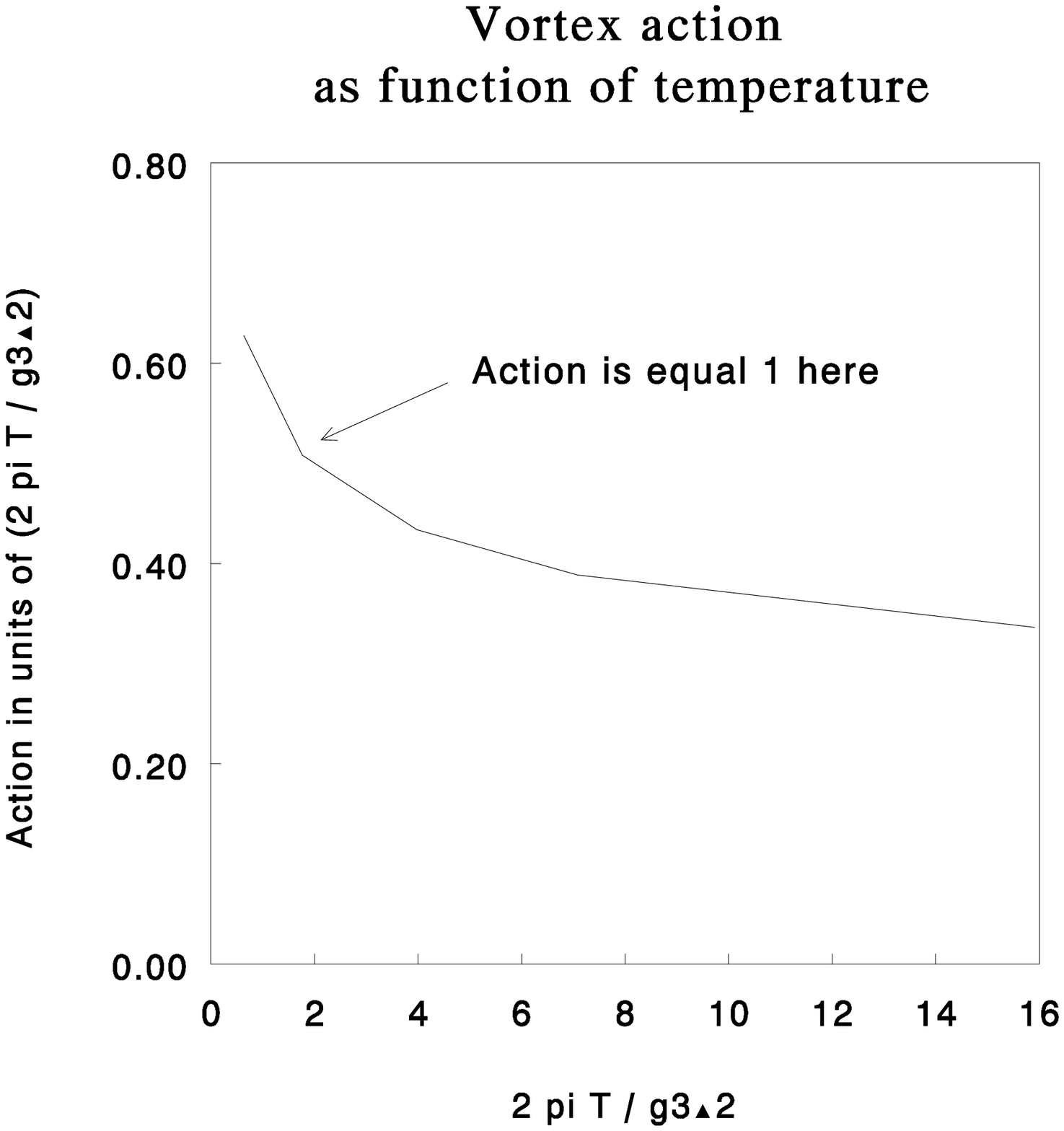}}
\caption[]{{\em The vortex action as function of
dimensionless temperature $T/g_3^2$}.}
\end{figure}

To build the $2+1$-dimensional vacuum at high temperatures out of the
vortices, one needs first of all to gauge rotate the `Higgs field'
$A_0^a$ \ur{A0} to, say, the third direction in colour space: one
cannot add up vortices with different colour orientations of $A_0$ at
infinity. This is achieved with a help of a $\phi$-dependent gauge
transformation

\beq
A_\mu^\prime = U^\dagger(\phi)A_\mu U(\phi)+i\delta_{\mu\phi}
\frac{1}{\rho}U^\dagger(\phi)\frac{d}{d\phi}U(\phi),\qquad
U(\phi)=\frac{1}{\surd 2}\left(\begin{array}{cc}
e^{i\phi/2} & ie^{i\phi/2} \\
e^{-i\phi/2} & -ie^{-i\phi/2}\end{array}\right).
\la{gt}\eeq
Under this gauge transformation the $A_0$ component of the YM field
becomes proportional to $\tau_3$,
$A_0^{\prime\,a}=\delta^{a3}\,2\pi T\,\nu(\rho)$. The gauge
transformation \ur{gt} is, however, discontinuous at $\phi=2\pi$,
therefore the azimuthal component $A_\phi^\prime$ will now have
a `Dirac-surface' singularity at $\phi=0$:

\beq
A_\phi^{\prime\,a}=\delta^{a2}\frac{1}{\rho}
\left[1-\mu(\rho)-2\pi\delta(\phi)\right].
\la{Aphiprime}\eeq
The Wilson loop \ur{WJ} is, naturally, preserved by this gauge
transformation: it remains $(-1)^{2J}$ at $\rho\to\infty$ (for the
representation labelled by spin $J$).

Having oriented the `Higgs expectation value' $A_0(\infty)$ in one
colour direction it is now possible to add up many vortices, each of
them necessarily carrying a singular Dirac surface. Apart from the
factor $\exp(-S_{{\rm vortex}})$ the statistical weight of a vortex
is determined by the fluctuation determinant, in particular by the
zero modes of the solution. We expect three zero modes here, two
of which are associated with the spatial position of the vortex center
$z_i$, and one pure gauge cyclic mode associated with shifts in the
`time' direction. (The latter can be revealed if one uses a gauge with
$A_0$ explicitly dependent on time.)

Denoting the (uncalculated) prefactor arising from the fluctuation
determinant and from zero modes by $\kappa(T,g_3)$ we can write
the vortex partition function as that of a gas,

\beq
{\cal Z}_{{\rm vortex}}=\sum_N\frac{1}{N!}
\left(\int\! d^2z\,\kappa\,\exp(-S_{{\rm vortex}})\right)^N,
\la{partfu}\eeq
implying that at large $T$ vortices are dilute and thus neglecting
their interactions. It follows immediatelly from \eq{partfu} that
the spatial density of vortices (i.e. number per unit area) is

\beq
n=\kappa\exp(-S_{{\rm vortex}}).
\la{dens}\eeq
At large $T$ the numbers of noninteracting vortices inside any large
area are Poisson-distributed. Therefore, the average of large
Wilson loops can be calculated as

\beq
\langle W_J\rangle =\sum_N \left[(-1)^{2J}\right]^N
\frac{(n\cdot{\rm Area})^N}{N!}e^{-n\cdot{\rm Area}}
=\left\{\begin{array}{cc}e^{-2n\cdot {\rm Area}}&
{\rm at}\;J={\rm half}\!\!-\!\!{\rm integer},\\
1 & {\rm at}\;J={\rm integer},\end{array}\right.
\la{WJav}\eeq
so that the string tension for half-integer representations is
twice the vortex density \cite{G1}, and zero for integer
represenations.

It should be recalled, however, that at $T\to\infty$ the $2+1$
dimensional system reduces to two dimensions with a coupling constant
$g_2^2=g_3^2T$, and in two dimensions there is a trivial perturbative
confinement with a nonzero string tension in any representation,

\beq
\sigma_2=g_2^2\,J(J+1).
\la{sigma2d}\eeq
The above vortex-induced string tension should be thus regarded as
an exponentially small addition to \eq{sigma2d}: it leads to a
deviation from the `Casimir scaling' of \eq{sigma2d}. It would be
instructive to study the high-temperature $2+1$ dimensional YM theory
in lattice simulations, if only because one can learn to identify
physical vortices by comparing objects found from maximal center gauge
fixing with the continuum profiles presented here. Such experience
might be useful in higher dimensions.

As one lowers T the vortex action decreases; at $T\approx g_3^2$
the action becomes of the order of unity, therefore, the vortices are
not suppressed anymore. Unfortunately, the quantitative theory fails
at this point: first, because the effective action \ur{Eperp1} is not
accurate anymore, second, because the vortices become long and start to
bend, third, because one can hardly neglect interactions between dense
vortices. Much more serious efforts are needed to describe vortices in
this case (if they exist). Nevertheless, to our mind it is useful
to know that at least in the limit of high temperatures vortices do
exist in a pure Yang--Mills theory, and their basic properties are
established.  \\

Useful discussions with Jeff Greensite, Victor Petrov and Gerard 't Hooft
are gratefully acknowledged.

\end{document}